\documentclass[aps,prl,preprint,groupedaddress,12pt]{revtex4-2}

\usepackage{array}
\usepackage{multirow}
\usepackage{float}
\usepackage{tikz}

\usepackage{amsmath}
\usepackage{amssymb}
\usepackage{graphicx}
\usepackage{color, colortbl}

\definecolor{LightCyan}{rgb}{0.88,1,1}
\definecolor{LightCyan1}{rgb}{0.8,1,1}
\definecolor{LightCyan2}{rgb}{0,0.86,1}
\definecolor{LightCyan3}{rgb}{0.7,0.9,1}

\begin{document}

\title{Universal limiting transition temperature for the high $T_\mathrm{c}$ superconductors}
\author{Moon-Sun Nam}
\address{The Clarendon Laboratory, Department of Physics, University of Oxford, Oxford OX1 3PU, UK}
\author{Arzhang Ardavan}
\address{The Clarendon Laboratory, Department of Physics, University of Oxford, Oxford OX1 3PU, UK}

\date{\today}

\begin{abstract}
Since their discovery three decades ago it has emerged that the physics of high-$T_\mathrm{c}$ cuprate superconductors is characterised by multiple temperature scales, and a phenomenology that deviates significantly from the conventional para\-digm of superconductivity~\cite{Lee,Keimer-review,Tinkham}. Below  the ``pseudogap'' temperature, $T^*$,  a range of experiments indicate a reduction in the density of states \cite{Timusk}. Lower in temperature, $T_\mathrm{c}$ marks the onset of a bulk zero-resistance state. Intermediate between these, numerous other temperature boundaries and cross-overs have been identified, often postulated to be associated with fluctuations of some kind~\cite{Lee,Keimer-review}. However, for the most part there is little consensus either over their definitions or the physical mechanisms at play. One of these temperature scales is the Nernst onset temperature, $T_\mathrm{onset}$, below which thermoelectric phenomena characteristic of superconductivity are observed. Depending on the material, the difference between $T_\mathrm{onset}$ and $T_\mathrm{c}$ ranges from almost nothing to over 100~K. 
In this paper,
we identify $T_\mathrm{onset}$ from published experimental data according to a consistent definition; this reveals a remarkable consistency of behaviour across the whole high-$T_\mathrm{c}$ family, despite the appreciable  variations in other parameters. Our analysis suggests that in the cuprates there is an inherent limit to $T_\mathrm{c} \sim 135$~K.
We compare this behaviour with other strongly-correlated superconductors and propose a unified picture of superconducting fluctuations close to the Mott state.
\end{abstract}

\maketitle

The Nernst effect, the transverse electric field  $E_{y}$  generated by crossed temperature gradient $\nabla_x T$ and magnetic field $B_{z}$,  
\begin{equation}
E_y = -e_N \nabla_x T = -N B_z \nabla_x T 
\end{equation}
where $e_N$ is the Nernst signal and $N$ is the Nernst coefficient. In the normal state, 
$N = N_{\mathrm n} = -\frac{\pi^{2}}{3} \frac{k_{B}^{2}T}{m} \left. \frac{\partial \tau}{\partial \epsilon} \right\vert_{\epsilon _{F}}$ 
is usually expected to be small (though its magnitude can be enhanced under some circumstances~\cite{Behnia1,Behnia2,LSCO-Nd/Eu,YBCO-Taillefer1,YBCO-Taillefer2}), magnetic field independent for small fields, and linear in temperature~\cite{Sondheimer}. 

A qualitatively larger signal is expected in a superconducting vortex liquid state, because the flow of heat-carrying vortices down a temperature gradient generates a transverse phase-slip. An appreciable Nernst coefficient $N=N_{\mathrm s}+N_{\mathrm n}$ is measured above $T_\mathrm{c}$ in many high-$T_\mathrm{c}$ materials~\cite{Behnia2} and some other strongly correlated superconductors~\cite{Behnia1}. The extra contribution $N_{\mathrm s}$ has been interpreted by some as indicating the presence of a phase fluctuating superconducting state; in this state, the superconducting phase is postulated to be sufficiently coherent to support vortices (i.e., presumably, the phase correlation-length exceeds the superconducting coherence length defining the dimensions of a vortex core), but does not exhibit the long-range order required for a bulk zero-resistance state~\cite{Emery-nature,Ong-nernst,LSCO-Ong-prb}. Others have argued that Gaussian fluctuations of the amplitude of the superconducting order parameter should also lead to an appreciable Nernst signal~\cite{Ullah,ush-prl-2002}, and there has been some debate over whether the Nernst effect observed above $T_\mathrm{c}$ arises from phase- or amplitude-fluctuations of the order parameter. In reality it seems likely that both fluctuation mechanisms are at play in different materials and different temperature regimes, but the consensus has emerged that the significant Nernst contribution $N_{\mathrm s}$ is characteristic of superconducting fluctuations.
In contrast to $N_\mathrm{n}$, $N_\mathrm{s}$ is expected to increase rapidly as temperature decreases, and to develop a magnetic field dependence.

Here, we have reviewed the extensive literature on the Nernst effect in high-$T_\mathrm{c}$ superconductors, and we have extracted $T_\mathrm{onset}$ from all high-$T_\mathrm{c}$ families for which data are reported.
We have attempted to use a consistent definition, where $T_\mathrm{onset}$ is identified as the temperature at which there is a deviation of $N$ from linearity in temperature. When the magnetic field dependence is also reported, we find that in most cases, this definition of $T_\mathrm{onset}$ is coincident with the temperature at which a magnetic field dependence emerges in $N$. We note in Table 1 which criterion we used to extract $T_\mathrm{onset}$. Most values for $T_\mathrm{onset}$ that we extract are in a good agreement with the original identifications~\cite{onset-definition}. 

In Fig.~1 we plot the ratio of $T_\mathrm{onset}/ T_\mathrm{c}$ against $T_\mathrm{c}$. A striking trends emerges: $T_\mathrm{onset}/T_\mathrm{c}$ is clustered strongly along a line $T_\mathrm{onset}/ T_\mathrm{c} \sim T_\mathrm{c}^{-0.79} $. 
The common behaviour across this broad range of materials is striking.
If $T_\mathrm{onset}$ marks the temperature below which there are significant superconducting fluctuations, a picture emerges of a class of materials in which what varies is how effectively they realise the ``potential'' for a phase-coherent superconducting state. Materials lying to the left of Fig.~1, and exhibiting a large $T_\mathrm{onset}/T_\mathrm{c}$ ratio, fail to stabilise a coherent state until temperatures well below the onset of fluctuations.  
In contrast, materials to the right of the plot, typically the more three-dimensional structures with multiple of Cu-O layers, establish the long-range superconducting order at the temperature closer to the temperature at which fluctuations become significant.  

The materials exhibiting the larger values of $T_\mathrm{onset}/T_\mathrm{c}$ are generally the more two-dimensional, under-doped, single layered cuprates, and this limit is most naturally described in a phase-fluctuating picture~\cite{Emery-nature,Ong-nernst,LSCO-Ong-prb,LSCO-Ong-nature}.  However, this description is less applicable to the more three-dimensional and optimally- or over-doped materials, where an amplitude-fluctuating scenario could be more relevant, and is partially successful in describing the behaviour of optimally-doped materials~\cite{Ullah,ush-prl-2002}. It is likely that a ``hybrid'' description is necessary for a quantitive account of the fluctuating state across the range of Fig.~1, but to our knowledge such a description does not yet exist.

We note that the trend-line for $T_\mathrm{onset}/T_\mathrm{c}$ crosses 1 at about 135 K at the upper limit, the maximum ambient-pressure transition temperature found experimentally more than two decades ago in Hg-based superconductors\,(HgBa$_{2}$Ca$_{2}$Cu$_{3}$O$_{1+x}$ with $T_\mathrm{c} \sim 133$ K)~\cite{Hg1223}. 
This supports the picture that across the families of cuprate superconductors there is some kind of intrinsic temperature scale (of the order 100~K) representing the ``potential'' for superconductivity, and that this potential is fully realised under certain circumstances of doping and crystal structure or dimensionality. This would also argue against the possibility that the cuprates are a good system in which to search for superconductors with $T_\mathrm{c}$ exceeding this intrinsic scale (an observation that is borne out by the history of the evolution of the maximum $T_\mathrm{c}$ in these materials). 

Fig.~2 shows how $T_\mathrm{onset}$  varies with hole concentration $p$, for all samples plotted in Fig.~1 for which doping data are available. $T_\mathrm{onset}$ is clustered between 75~K and 125~K for most materials, a much narrower range than the corresponding range of $T_\mathrm{c}$.   
The dashed line in Fig.~2 shows the ambient-pressure superconducting phase boundary as a function of $p$ for the highest $T_\mathrm{c}$ Hg-based cuprate superconductors. The striking observation is that the values of $T_\mathrm{onset}$ for a very wide range of materials and dopings are clustered around the maximum $T_\mathrm{c}$ found in any ambient-pressure cuprate superconductor.

Within each family of materials, we find that $T_\mathrm{onset}$ has a weak dependence on $p$ (see Fig.~2 inset showing this for LSCO), exhibiting a maximum at $p \sim 0.125$, whereas $T_\mathrm{c}$ is maximum at $p \sim 0.16$.
The occurrence of the peak in $T_\mathrm{onset}$ at lower $p$ than the peak in $T_\mathrm{c}$ suggests that the physical mechanisms determining $T_\mathrm{onset}$ are shared with those determining both $T^*$ and $T_\mathrm{c}$. Thus, the enhancement of $T^*$ below optimum $p$ acts to enhance $T_\mathrm{onset}$, while at lower $p$, $T_\mathrm{c}$ is suppressed significantly, and, since $T_\mathrm{onset}$ is ultimately related to the superconductivity, $T_\mathrm{onset}$ is also eventually suppressed as $p$ decreases. The fact that $T_\mathrm{onset}$ seems to be connected to both $T^*$ and $T_\mathrm{c}$ for the entire family of materials suggests that the pseudogap is involved in the superconductivity. (We note in passing that the peak in $T_\mathrm{onset}$ is approximately coincident with the doping at which charge order is most stable~\cite{Kivelson1,Kivelson2,S.Davis,Lee2}).  

In summary, we have identified a consistent behaviour in the Nernst effect across a wide range of high-$T_\mathrm{c}$ superconductors. First, we find that $T_\mathrm{onset}/T_\mathrm{c}$ follows a consistent trend 
across the family and that $T_\mathrm{onset}$ is  limited at around 135K, close to the highest ambient-pressure $T_\mathrm{c}$ observed in the cuprate superconductors.  
Second, the $p$ dependence of  $T_\mathrm{onset}$  across each of the high-$T_\mathrm{c}$  families indicates that the superconducting fluctuation temperature scale peaks at $p \sim$ 0.125, lower than the doping optimising $T_\mathrm{c}$, indicating an interplay between the two temperature scales $T_\mathrm{c}$ and $T^{\ast}$.

Broadly, $T_\mathrm{onset}/T_\mathrm{c}$ is larger for materials with lower doping (i.e., closer to the Mott state) and in which doping-dependent charge order occurs. The phase coherence of a superconducting state is associated with uncertainty in particle number~\cite{Tinkham}, while in the Mott and charge-ordered states (which are stabilised by Coulomb interactions), number-uncertainty incurs an energy penalty. Thus proximity to a Mott or charge-ordered state might be expected to enhance phase fluctuations in the superconducting state.  

Of course, the cuprates are not the only materials to exhibit superconductivity in the proximity of a Mott or charge-ordered state. The $\kappa$-phase quasi-two-dimensional BEDT-TTF organic molecular metals~\cite{Ishiguro} offer a complementary family of strongly-correlated superconductors, in which the Coulomb energy scale, $U$, is set largely by the dimerised packing of the BEDT-TTF molecules. The hybridisation between the dimer orbitals sets the band width energy scale, $t$ (see Fig.~3), and this quantity is tuneable by hydrostatic or chemical pressure, yielding a phase diagram in which a Mott state can be tuned into a strongly-correlated superconductor as a function of $t/U$ (see Fig.~3). With further (hydrostatic or chemical) pressure, the correlation effects can be reduced further, and, eventually, the superconductivity is completely suppressed.

Throughout the $\kappa$-phase series the stoichiometry is preserved, 
so this family of materials gives us the opportunity to explore the phase diagram along an axis orthogonal to that relevant for the cuprates (see Fig.~3). In particular, Nernst measurements on these materials reveal a fluctuating superconducting state neighbouring the Mott state along the $t/U$ axis~\cite{nam2007,nam2013} (compounds $\kappa$-(BEDT-TTF)$_{2}$Cu[N(CN)$_{2}$]]Cl$_{1-x}$Br$_{x}$ and  $\kappa$-(BEDT-TTF)$_{2}$Cu[N(CN)$_{2}$]Br); further increasing $t/U$ seems to suppress the fluctuating state, yielding a system in which there is no sign of fluctuations above $T_\mathrm{c}$ ($ \kappa$-(BEDT-TTF)-Cu(NCS)$_{2}$).

The organics exhibit an equivalent trend in $T_\mathrm{onset}/T_\mathrm{c}$ to what we have identified above in the cuprates in Fig.~1. There is much less Nernst data published on organics, but what is available is plotted in Fig.~3. However, whereas in the cuprates $T_\mathrm{onset}$ is rather uniform, in the $\kappa$-(BEDT-TTF) materials $T_\mathrm{onset}$ varies greatly. It is therefore all the more remarkable that the plot of $T_\mathrm{onset}/T_\mathrm{c}$ nevertheless seems to follow a trend analogous that exhibited in the cuprates, $T_\mathrm{onset}/ T_\mathrm{c} \sim T_\mathrm{c}^{-0.79} $ with maximum possible $T_\mathrm{c} \sim 12$~K

This observation motivates the hypothesis that the unifying $t/U$ vs doping phase diagram is as shown in Fig.~4. Here, we have normalised the vertical (temperature axis) to the Mott ordering temperature at zero doping (cuprates) and at minimum bandwidth (organics). Experiments show that there is a fluctuating superconducting state neighbouring the Mott state along both the $t/U$ axis and the doping axis. The argument that proximity to the Mott state enhances superconducting phase fluctuations should apply throughout, so we suggest that the fluctuating state in organics is connected to the fluctuating state in cuprates. As yet, we do not know of an experimental system in which we can probe this bridging regime. However, recent experiments on films on $\kappa$-BEDT-TTF materials have allowed both doping (by gating the sample) and the application of strain~\cite{Doping_organics1,Doping_organics2}, offering the tantalising possibility of examining the fluctuating superconducting state as a continuous function of both $t/U$ and doping.

We also note that in the cuprates literature there is significant variation in the reported temperature scales associated with phenomena such as charge order and fluctuation Nernst effect. Both are expected to be sensitive to magnetic field, which suppresses superconductivity but stabilises charge order~\cite{FICDW}. The wide range of fields at which the reported experiments are conducted may therefore be masking trends; there would be value in experiments probing these temperature scales conducted at uniform (low) magnetic fields.

\newpage

\pagestyle{empty}


\newcolumntype{g}{>{\columncolor{LightCyan}}c}
\newcolumntype{f}{>{\columncolor{LightCyan2}}c}
\newcolumntype{e}{>{\columncolor{LightCyan3}}c}
\setlength{\tabcolsep}{4pt}
\renewcommand{\arraystretch}{0.4}
\begin{table}[h]
\vspace{-0.5cm}
\footnotesize
\centering
\begin{tabular}{|e|c|g|g|g|c|c|c|}\hline
\rowcolor{LightCyan2}
  						  			& $x$  & $p$    &T$_{c}$ (K)  		&T$_{onset}$ (K)		& T- linear & B-dep. & ref.  \\  \hline 
  La$ _{2-x}$Sr$_{x}$CuO$_{4}$ UD  		& 0.03   & 0.03   & 0   	  		& none    			&   $\circ$    &   	& \cite{Ong-nernst}\\
  (LSCO-UD)									& 0.05   & 0.05   & $<$4 		&  $40 \pm10$   	& $\circ$      &   	&  \cite{Ong-nernst} \\
   									& 0.06   & 0.06   & 8   	  		&    $30 \pm10$   	&   $\circ$    &   	& \cite{LSCO-Behnia}\\
   									& 0.07   &  0.07  &  11     	 	&    $95 \pm10$   	&  $\circ$     &   	&  \cite{Ong-nernst,LSCO-Ong-prb}\\
  									& 0.10   & 0.10   & 25       		&  $125 \pm10$     & $\circ$      &$\circ$ &  \cite{Ong-nernst,LSCO-Ong-nature}\\
  									& 0.12   & 0.12   &  29      		&     $115 \pm10$  	&  $\circ$     &   	&  \cite{Ong-nernst}\\  \hline
  La$ _{2-x}$Sr$_{x}$CuO$_{4}$ OD 		& 0.17   &  0.17  &   36     		&    $90 \pm10$   	&  $\circ$     &   	&   \cite{Ong-nernst,LSCO-Ong-prb}\\	       (LSCO-OD)									& 0.20   &  0.20  &   28     		&   $75 \pm10$    	& $\circ$      &   	&   \cite{Ong-nernst,LSCO-Ong-prb}\\   \hline
   La$ _{1.8-x}$Eu$_{0.2}$ Sr$_{x}$CuO$_{4}$ & 0.125  & 0.125 &   $5 \pm2$       &    $34 \pm4$        &      	   &$\circ$ &   \cite{LSCO-Nd/Eu}\\ 
		(Eu-LSCO)						       & 0.125  & 0.125  &   $5 \pm2$       &    $140 \pm10$    & $\circ$     &            &   \cite{LSCO-Nd/Eu}\\
 								       & 0.16    & 0.16    &   $16 \pm3$     &    $120 \pm10$    & $\circ$     &            &   \cite{LSCO-Nd/Eu}\\ \hline

  La$ _{1.6-x}$Nd$_{0.4}$ Sr$_{x}$CuO$_{4}$ & 0.2     &  0.2     &  $20 \pm1$      & $32 \pm4$      	& 		   & $\circ$& \cite{LSCO-Nd/Eu}  \\
 			(Nd-LSCO)							& 0.2     &  0.2    &  $20 \pm1$      & $70 \pm4$           &$\circ$  	   &   	        &  \cite{LSCO-Nd/Eu} \\ \hline
   La$ _{2-x}$Ba$_{x}$CuO$_{4}$ (LBaSCO)		        & 0.125 &  0.125 &   5  		       & $110 \pm10$        & $\circ$     &   		& \cite{LBaCuO} \\ \hline
  Bi$_{2}$Sr$_{2-x}$La$_{x}$CuO$_{6}$ UD    &  0.74 &   0.109 & 21    		       &  $72 \pm 5$          &   $\circ$   &   		& \cite{Bi2201-Eu}\\
     	(La-Bi2201-UD)							        &  0.73 &  0.11    &  18                   &    $75 \pm5$         &    $\circ$  &           & \cite{Bi2201-Eu} \\
     								        & 0.6    &  0.13    &   17                 &   $100 \pm10$       &   $\circ$   &           &  \cite{LSCO-Ong-prb}\\
     								        &0.55   &  0.135  &  31   	       &    $81 \pm5$          &      	    & $\circ$&  \cite{Bi2201-Eu}\\
       						 		        & 0.5    &  0.14    &   28  	       &   $80 \pm10$         &   $\circ$   &              & \cite{LSCO-Ong-prb} \\ \hline
  Bi$_{2}$Sr$_{2-x}$La$_{x}$CuO$_{6}$ OP     &  0.4  &  0.16    &  28   		       &    $70 \pm10$        &      	    &  $\circ$  &  \cite{LSCO-Ong-prb} \\
   		(La-Bi2201-OP)								& 0.4   &  0.16    & 27    		       & $68 \pm5$             &   $\circ$   &               &  \cite{Bi2201-Eu} \\
     								        & 0.38  &  0.162  &  32                  &  $64 \pm5$             &    $\circ$ &               & \cite{Bi2201-Eu}  \\ \hline
  Bi$_{2}$Sr$_{2-x}$La$_{x}$CuO$_{6}$ OD (La-Bi2201-OD)	   & 0.2    & 0.18     & 24    		       & $59 \pm10$            &  $\circ$   &   	     & \cite{Bi2201-Eu} \\ \hline
  Bi$_{2}$Sr$_{2-x}$Eu$_{x}$CuO$_{6}$ UD    & 0.40  &  0.13    &  16                  &  $71 \pm10$            &   $\circ$   &             &\cite{Bi2201-Eu}  \\
   		(Eu-Bi2201-UD)							        & 0.42  &  0.10    &  12                  &  $69 \pm10$            &   $\circ$   &              &  \cite{Bi2201-Eu}\\ \hline
  Bi$_{2}$Sr$_{2-x}$Eu$_{x}$CuO$_{6}$ OP (Eu-Bi2201-OP)	   & 0.28  & 0.16     &  17    	       &  $67 \pm10$            & $\circ$     &              & \cite{Bi2201-Eu}  \\ \hline
  Bi$_{2}$Sr$_{2}$CaCu$_{2}$O$_{8 + \delta}$ UD &   & 0.09     & 50     	       &   $118 \pm5$           &                 &$\circ$   & \cite{Ong-nernst, Bi2212} \\
                  (Bi2212-UD)                                                                 &   & 0.12     & 75                   &    $130 \pm10$       &  $\circ$     &             & \cite{Ong-nernst, Bi2212}  \\  \hline
  Bi$_{2}$Sr$_{2}$CaCu$_{2}$O$_{8 + \delta}$ OP (Bi2212-OP)  &   &  0.16    & 91    	        &  $125 \pm5$          &                  &  $\circ$  & \cite{Ong-nernst, Bi2212} \\   \hline
  Bi$_{2}$Sr$_{2}$CaCu$_{2}$O$_{8 + \delta}$ OD  &   &  0.202  &  77     	        & $100 \pm5$           &   $\circ$     &             & \cite{Ong-nernst, Bi2212}  \\ 
                   (Bi2212-OD)                                                                &   & 0.219   &   65    	        &   $87 \pm10$          &  $\circ$     &             & \cite{Ong-nernst, Bi2212}  \\  \hline
  YBa$_{2}$Cu$_{3}$O$_{y}$ UD 			  &  6.5  & 0.105     & 50     		& $120 \pm10$         &   $\circ$     &             &  \cite{Ong-nernst} \\
     (YBCO-UD)		  					  & 6.54 &  0.11    &  61.5                  & $85 \pm5$             &                   & $\circ$  &  \cite{YBCO-Taillefer2} \\
               		  					          & 6.67 &  0.12    &  66                  & $87.5 \pm5$             &                   & $\circ$  &  \cite{YBCO-Taillefer1,YBCO-Taillefer2} \\
				  					  & 6.75  &  0.132   & 75                   & $100 \pm5$              &      	        &$\circ$   &  \cite{YBCO-Taillefer2} \\ 
									  & 6.86  &  0.15   & 91                   & $106 \pm5$              &      	        &$\circ$   &  \cite{YBCO-Taillefer2} \\  \hline
  YBa$_{2}$Cu$_{3}$O$_{y}$ OD (YBCO-OD)			  &  6.99 &  0.18   &  92                  & $105 \pm10$          &                  &  $\circ$   &   \cite{Ong-nernst}  \\ \hline
  YBa$_{2}$Cu$_{3}$O$_{6.6}$ damaged UD     & 6.6    &           &57      		&  $85 \pm2$    	    &      	       & $\circ$   &  \cite{YBCO-damaged} \\
    	(YBCO-damaged UD)			             & 6.6    &            & 45.1     		&   $83 \pm2$            &  $\circ$     &              &  \cite{YBCO-damaged} \\
 				                                            & 6.6    &            & 24.2     		&   74                         &   $\circ$    &              &   \cite{YBCO-damaged} \\
 				                                            & 6.6    &            &3      		&  70                           &  $\circ$     &              &   \cite{YBCO-damaged} \\ \hline
  YBa$_{2}$Cu$_{3}$O$_{7.0}$ damaged OP 	  & 7.0    &          &92.6      		&  $103 \pm2$    	    &  $\circ$      &               &  \cite{YBCO-damaged} \\
    			(YBCO-damaged OP)	                   &7.0    &            & 79.5     		&   $97\pm2$            &  $\circ$     &              &  \cite{YBCO-damaged} \\
 				                                            & 7.0    &            & 48.6     		&   82                         &   $\circ$    &              &   \cite{YBCO-damaged} \\  \hline
   YBa$_{2}$(Cu$_{1-x}$Zn$_{x}$)$_{3}$O$_{7- \delta}$ film  &  0       &           & 90      &  $105 \pm5$     & $\circ$     &   		& \cite{YBCO-Zn-film} \\
    		         	(Zn-YBCO film) 							      & 0.005 &           & 84      & $97\pm5$         &  $\circ$    &   		& \cite{YBCO-Zn-film} \\
		          							              &  0.01   &          & 79      &   $89 \pm5$       &     	        &  $\circ$ & \cite{YBCO-Zn-film} \\
     			          							      & 0.02    &          & 67      &  $75\pm5$         &  $\circ$    &  		 &  \cite{YBCO-Zn-film} \\  \hline
  Y$_{1-x}$Pr$_{x}$Ba$_{2}$Cu$_{3}$O$_{7- \delta}$ film	     &  0         & 0.19  & 90      &  $105 \pm5$        & $\circ$     &               & \cite{YBCO-Pr-film} \\
   			       (Pr-YBCO film)    							     & 0.1       &          &  83     & $100 \pm5$         & $\circ$     &               &  \cite{YBCO-Pr-film}  \\
    			          							     & 0.2       &          & 65       &  $95 \pm5$          &                 &$\circ$    &  \cite{YBCO-Pr-film}  \\
    		          								     & 0.3       &          & 50       &  $85 \pm5$          &   $\circ$   &               &  \cite{YBCO-Pr-film}  \\
      		         								     & 0.4       &          &  40       &   $80 \pm5$         &    $\circ$  &               &  \cite{YBCO-Pr-film}  \\ \hline
  Y$_{0.9}$Ca$_{0.1}$Ba$_{2}$Cu$_{3}$O$_{y}$ film	     &             & 0.12  &  81.7    &  $112 \pm5$      &  $\circ$    &   & \cite{YBCO-Ca-film}  \\
  			(Ca-YBCO film)           							     &             & 0.14  &  84.5    &  $115 \pm5$      &   $\circ$   &   & \cite{YBCO-Ca-film} \\
			          							     &             &  0.2   &  82.2    &   $120 \pm5$     &   $\circ$   &   & \cite{YBCO-Ca-film} \\  \hline
   Bi$_{2}$Sr$_{2}$Ca$_{2}$Cu$_{3}$O$_{10 + \delta}$ OP (Bi2223OP)    &   &0.16   &  109   &   $135 \pm5$    &      &  $\circ$  & \cite{Ong-nernst}  \\  \hline
\end{tabular}
\label{default}
\caption{High-$T_\mathrm{c}$ superconductors}
\end{table}

\newpage

\newcommand{\xdownarrow}[2][]{%
\left.{#1}\right\Downarrow{#2}}
  
\newcolumntype{g}{>{\columncolor{LightCyan}}c}
\newcolumntype{f}{>{\columncolor{LightCyan2}}c}
\newcolumntype{e}{>{\columncolor{LightCyan3}}c}
\setlength{\tabcolsep}{4pt}
\renewcommand{\arraystretch}{0.4}

\begin{table}[h]

\footnotesize
\centering

\begin{tabular}{|c||e|g|g|c|c|c|} \hline
\rowcolor{LightCyan2}

  			t/U		&	  			&T$_{c}$ (K)  		&T$_{onset}$ (K)		& T- linear & B-dep. & ref.  \\  \hline 				
Higher 	&$\kappa$-(BEDT-TTF)$_{2}$Cu(NCS)$_{2}$  & 9.8  			& 10 $\pm $1			& $\circ$ &  $\circ$  & \cite{nam2007}      \\
 \multirow{7}{*}{$\xdownarrow[\begin{gathered}
  \hfill \\
  \hfill \\
  \end{gathered}]{}$ }	
	&($ \kappa$-(BEDT-TTF)-Cu(NCS)$_{2}$ )			&				&					&		&		&     \\   \cline{2-7}  \hline
	&$\kappa$-(BEDT-TTF)$_{2}$Cu[N(CN)$_{2}$]Br & 11.6			&18 $\pm$ 2			 & $\circ$ &  $\circ$&  \cite{nam2007, nam2013}     \\ 
	&( $ \kappa$-(BEDT-TTF)-Br) 						&				&					&		&		& \\ \cline{2-7} \hline 
 	&$\kappa$-(BEDT-TTF)$_{2}$Cu[N(CN)$_{2}$]]Cl$_{0.2}$Br$_{0.8}$ & 11.8			&20 $\pm$ 2			 & $\circ$ &  $\circ$&  \cite{nam2013}      \\ 
	 &($ \kappa$-(BEDT-TTF)-Cl$_{0.2}$Br$_{0.8}$ )						&				&					&		&		& \\ \cline{2-7} \hline 
 	&$\kappa$-(BEDT-TTF)$_{2}$Cu[N(CN)$_{2}$]]Cl$_{0.27}$Br$_{0.73}$ & 10.9			&55 $\pm$ 5			 & $\circ$ &  $\circ$&  \cite{nam2013}     \\  
Lower & ($ \kappa$-(BEDT-TTF)-Cl$_{0.27}$Br$_{0.73}$ )						&				&					&		&		&  \\     \hline
\end{tabular}
\label{default}
\caption{$\kappa$-(BEDT-TTF) organic superconductors}
\end{table}

\newpage

\begin{figure}[h]
\includegraphics[height=11.5cm]{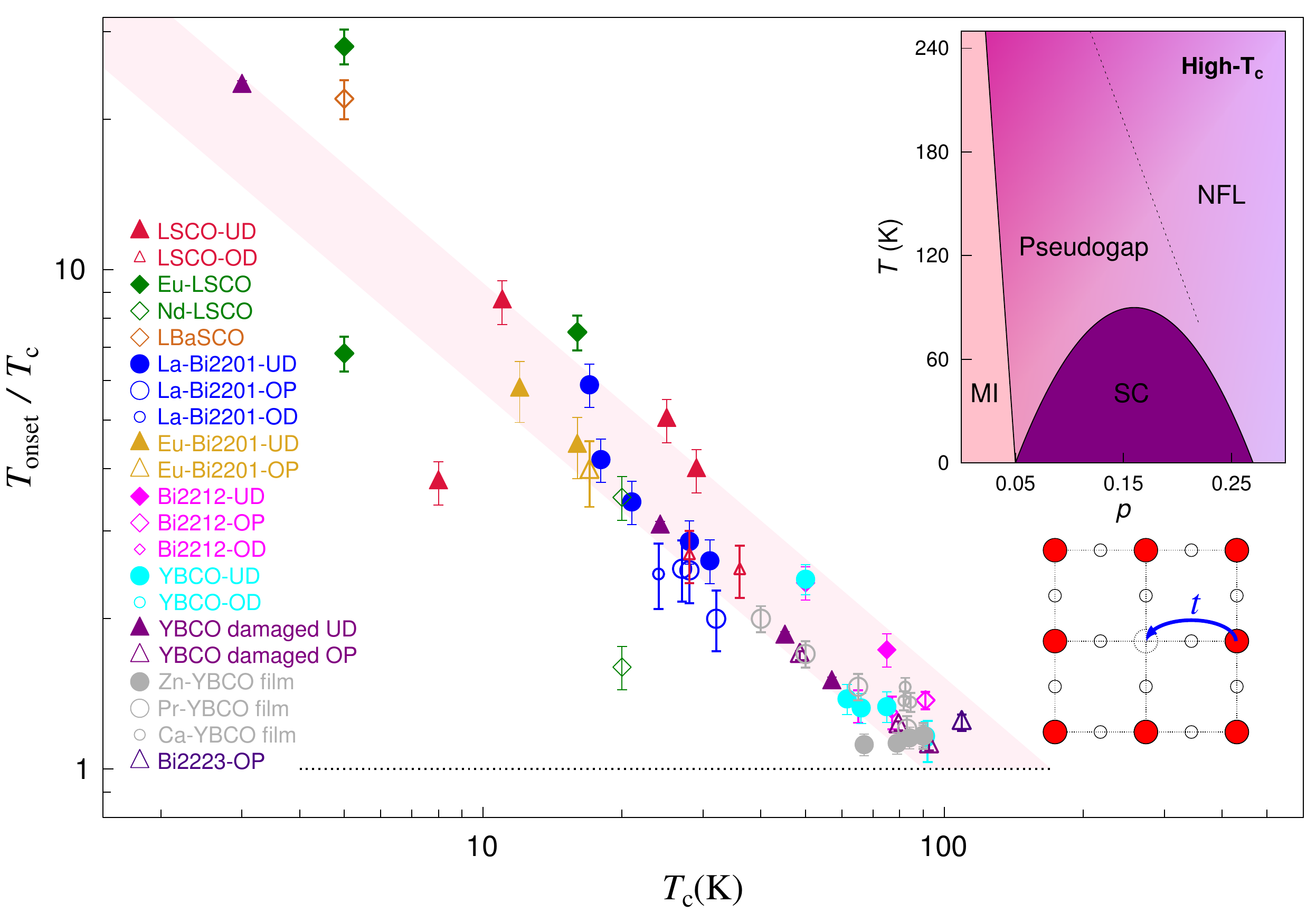}
\caption{Ratio of $T_\mathrm{onset}/ T_\mathrm{c}$ against $T_\mathrm{c}$ in high-$T_\mathrm{c}$ superconductors. The data are listed and referenced in Table~1 with references from which we extracted $T_\mathrm{onset}$. ``UD'' indicates under-doped, ``OP'' optimally-doped and ``OD'' over-doped. Inset: the phase diagram of the cuprates as a function of hole doping and temperature.}
\end{figure}
\newpage

\newpage 

\begin{figure}[h]
\includegraphics[height=11.5cm]{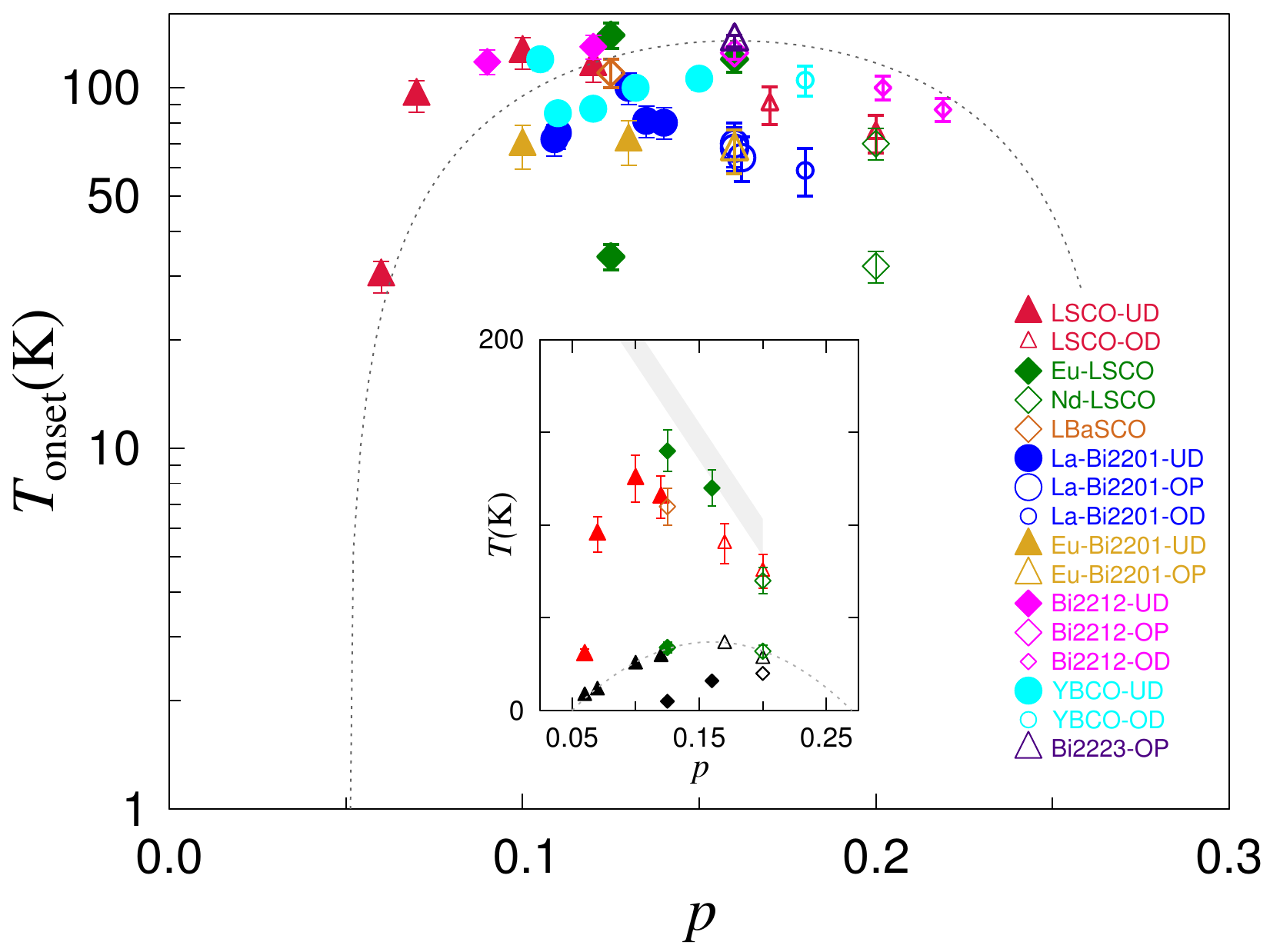}
\caption{$T_\mathrm{onset}$ as a function of hole concentration $p$. Inset: The temperatures  $T_\mathrm{c}$ (black symbols) and $T_\mathrm{onset}$ (coloured symbols) are plotted as a function of $p$ for a particular family, LSCO. The dotted line indicates  the long-range ordered superconducting phase boundary,
the grey shaded line indicates the pseudogap temperature $T^{\ast}$~\cite{Sawatzky}.}
\end{figure}

\newpage

\begin{figure}[h]
\includegraphics[height=11.5cm]{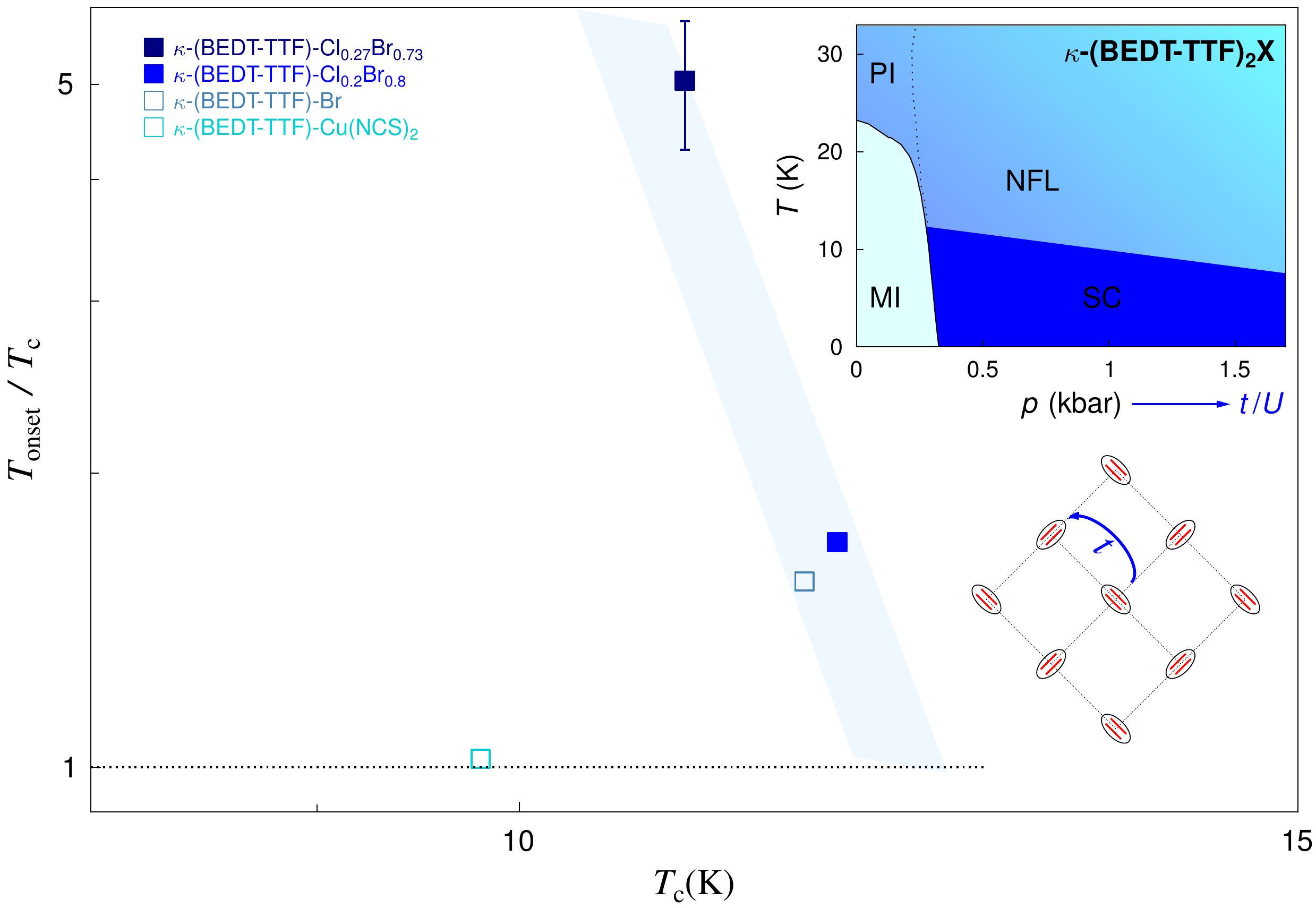}
\caption{Ratio of $T_\mathrm{onset}/ T_\mathrm{c}$ against $T_\mathrm{c}$ in $\kappa$-(BEDT-TTF) organic superconductors. The data are listed and referenced in Table~2 with references from which we extracted $T_\mathrm{onset}$.}
\end{figure}

\newpage

\begin{figure}[h]
\includegraphics[height=11.5cm]{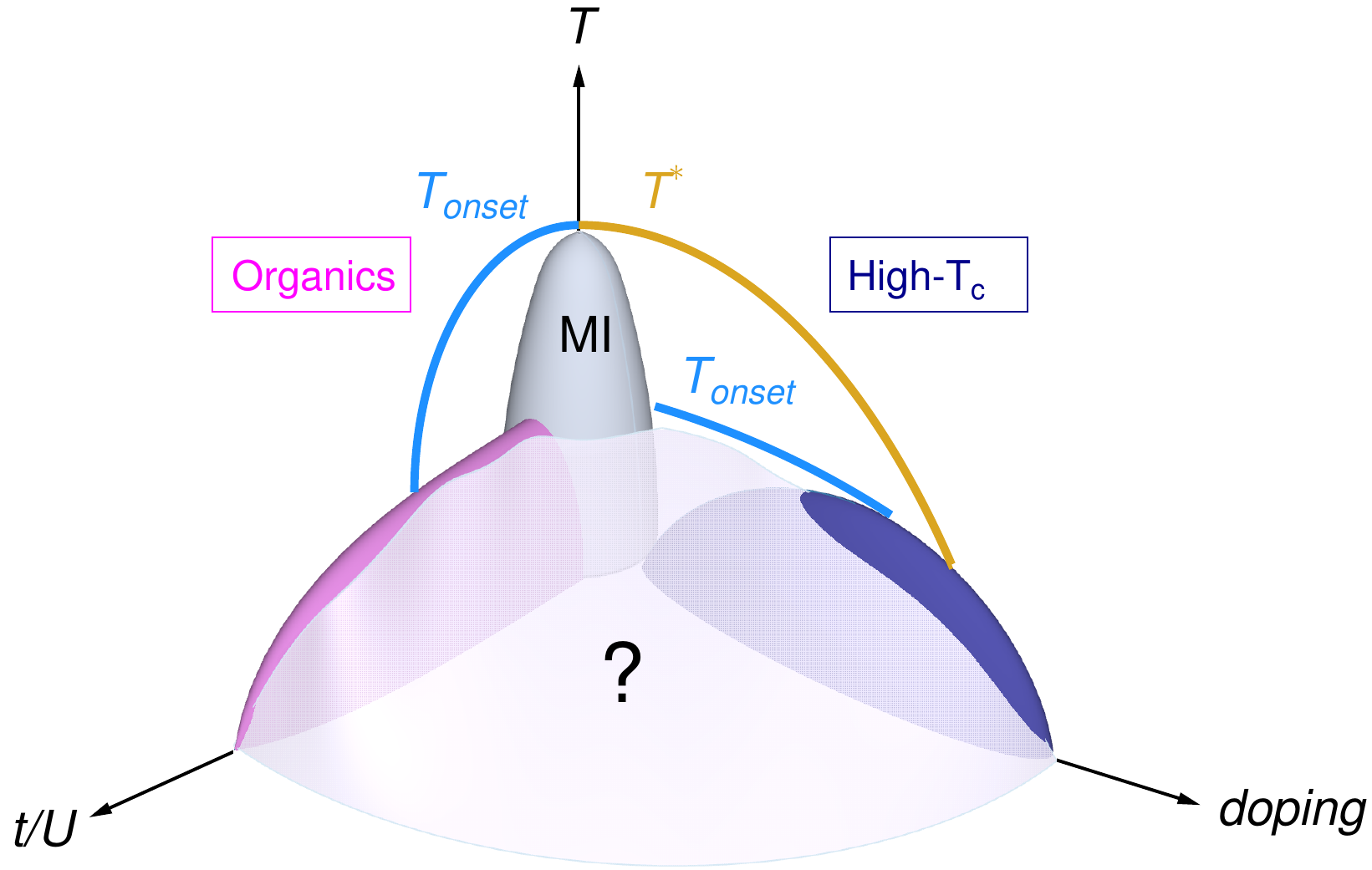}
\caption{Proposed unified $t/U$ vs doping phase diagram.}
\end{figure}
\newpage


\begin{thebibliography}{50}
\bibitem{onset-definition}
The advantage to identifying $T_\mathrm{onset}$ by the magnetic-field-dependent criterion is that it is absolutely empirical; it does not depend on any assumptions or models except the Gaussian fluctuations with 1/B dependency.  Unfortunately most studies do not report a magnetic field dependence.
\end{thebibliography}
\end{document}